\documentclass[12pt]{article}
\usepackage{hyperref}
\usepackage[english]{babel}
\usepackage{amsmath}
\usepackage{latexsym}
\usepackage{amssymb}
\usepackage[all]{xy}
\usepackage[dvips]{graphicx}
\usepackage{epsfig}
\usepackage{psfrag}

\numberwithin{equation}{section} \numberwithin{table}{section}
\numberwithin{figure}{section}

\begin{document}



\begin{titlepage}
   \begin{flushright}
{\small MPP-2013-157 }
  \end{flushright}

   \begin{center}

     \vspace{20mm}

     {\LARGE \bf Probing holographic semi-local quantum liquids with
     D-branes

     \vspace{3mm}}

     \vspace{10mm}

    Da-Wei Pang

     \vspace{5mm}

      {\small \sl Max-Planck-Institut f\"{u}r Physik (Werner-Heisenberg-Institut)\\
      F\"{o}hringer Ring 6, 80805 M\"{u}nchen, Germany}\\

     {\small \tt dwpang@mppmu.mpg.de}
     \vspace{10mm}

   \end{center}

\begin{abstract}
\baselineskip=18pt
We study dynamics of probe D-branes in $(d+2)$-dimensional background with general semi-locality. The background is characterized by a parameter $\eta$ and is conformal to $AdS_{2}\times\mathbb{R}^{d}$. We discuss thermodynamics of the probe D-branes and find that the entropy density is vanishing in the extremal limit, which indicates that the background may correspond to the true ground state of the system. We also clarify the conditions under which the specific heat matches to the behavior of a Fermi liquid or a Bose liquid. We calculate the current-current and density-density retarded Green's functions, from which we can obtain the AC conductivity and the zero sound mode. The AC conductivity scales as $\omega^{-1}$ when $d/\eta<2$ and $\omega^{-2\eta/d}$ when $d/\eta>2$, while it contains a logarithmic term in $\omega$ when $d/\eta=2$. We also observe that there is no quasi-particle excitation when $d/\eta\geq2$ and classify the conditions under which the quasi-particle description is valid when $d/\eta<2$.
\end{abstract}
\setcounter{page}{0}
\end{titlepage}

\pagestyle{plain} \baselineskip=19pt

\tableofcontents

\section{Introduction}
The AdS/CFT correspondence~\cite{Maldacena:1997re, Aharony:1999ti} has been proven to be a powerful tool for investigating dynamics of strongly-coupled field theories in the context of holography. Recently tremendous progress has been made in applying the AdS/CFT correspondence to condensed matter physics (AdS/CMT). Some widely studied systems in condensed matter physics, such as superconductors~\cite{Hartnoll:2008vx, Hartnoll:2008kx} and (non-)Fermi liquids~\cite{Liu:2009dm, Cubrovic:2009ye, Faulkner:2009wj}, have been well understood in the framework of AdS/CMT.

The simplest laboratory for studying AdS/CMT is the Reissner-Nordstr\"{o}m black hole (RN-AdS) in $d+2$-dimensional AdS spacetime. It was shown in~\cite{Liu:2009dm, Cubrovic:2009ye, Faulkner:2009wj} that this background provides holographic dual description of non-Fermi liquid, which can be seen by calculating the fermionic two-point function in this background. Such a property is closely related to the emergent $AdS_{2}$ geometry in the near horizon limit. Moreover, it was observed in~\cite{Iqbal:2011in} that although there is only nontrivial scaling in the time direction
due to $AdS_{2}$, the scaling dimension of a local operator and the corresponding correlation functions depend nontrivially on the momentum $k$. Therefore the phase described by $AdS_{2}\times\mathbb{R}^{d}$ is called a semi-local quantum liquid~\cite{Iqbal:2011in}.

However, one disadvantage of RN-AdS black hole is that there is a finite entropy at extremality, which is undesirable from a condensed matter point of view. One way to solve this problem is to consider black hole solutions in Einstein-Maxwell-dilaton theory~\cite{Charmousis:2010zz, Gouteraux:2011ce}, where the black hole entropy is vanishing at extremality. It should be emphasized that the Einstein-Maxwell-dilaton theory describes the IR physics of the corresponding system and therefore the exact solutions correspond to the near horizon geometry. It was pointed out in~\cite{Gouteraux:2011ce} that under certain conditions, the metric is conformal to $AdS_{2}\times\mathbb{R}^{d}$, which can be taken as a dual gravity solution of more general semi-local quantum liquids. Moreover, as pointed out in~\cite{Hartnoll:2012wm, Anantua:2012nj}, such a background can be obtained from the hyperscaling violation metric~\cite{Gouteraux:2011ce,Huijse:2011ef,Dong:2012hk}
\begin{equation}
ds^{2}=\frac{1}{r^{2}}\left(-\frac{dt^{2}}{r^{2d(z-1)/(d-\theta)}}+r^{2\theta/(d-\theta)}dr^{2}+\sum\limits_{i=1}^{d} dx_{i}^{2}\right),
\end{equation}
by taking the following limits:
\begin{equation}
\label{limit}
z\rightarrow\infty,~~\theta\rightarrow-\infty,~~\eta\equiv-\frac{\theta}{z}~~{\rm fixed},
\end{equation}
where $z$ denotes the dynamical exponent and $\theta$ is the hyperscaling violation parameter. For other references on holographic aspects of semi-local backgrounds, see~\cite{Donos:2012yi,Gouteraux:2012yr}.

One basic and important ingredient of realistic condensed matter systems is the presence of a finite density of charge carriers. In the holographic setup, the charge carriers can be modeled by probe D-branes~\cite{Hartnoll:2009ns}. It was shown in~\cite{Hartnoll:2009ns} that by putting probe D-branes in Lifshitz backgrounds, one can match the non-Fermi liquid scalings, such as linear resistivity, observed in strange metal regimes by choosing the dynamical critical exponent $z$ appropriately. For subsequent generalizations in this direction, see~\cite{sub}.

In this paper we consider dynamics of probe D-branes analytically in general $(d+2)$-dimensional semi-local backgrounds. Firstly we study thermodynamics of probe D-branes following~\cite{Karch:2007br} and observe that the resulting entropy density is vanishing in the extremal limit, which is different from the cases in which the background is Schwarzschild-AdS~\cite{Karch:2008fa} or Lifshitz black hole~\cite{HoyosBadajoz:2010kd}. We also clarify the conditions under which the specific heat of the system matches to the behavior of Fermi liquids or Bose liquids. Next we obtain the AC conductivity by evaluating the retarded Green's function using the matching method in~\cite{HoyosBadajoz:2010kd}. The AC conductivity exhibits different scaling powers in the frequency $\omega$, depending on the parameter $d/\eta$. Moreover, the AC conductivity contains a logarithmic term in $\omega$ when $d/\eta=2$, which is analogous to the $z=2$ case in~\cite{HoyosBadajoz:2010kd}. Finally we calculate the zero sound mode and classify the cases when the quasi-particle description applies. We find that when $d/\eta\geq2$, there is no quasi-particle description and for $d/\eta<2$, the existence of quasi-particles should be discussed case by case.

The rest of the paper is organized as follows: In section 2 we review the background with semi-locality in brief. Then we calculate the thermodynamical quantities, such as the energy density, entropy density and specific heat in section 3. In section 4 we compute the retarded Green's function by the matching technique, which enables us to read off the AC conductivity and the zero sound in section 5 and section 6 respectively. Finally a summary and discussion will be given in section 7.
\section{The background}
In this section we give a brief review on the background with semi-locality. Such a metric is an exact solution of the Einstein-Maxwell-dilaton theory:
\begin{equation}
 S=\int d^{d+2}x\sqrt{-g}\left(\frac{1}{2\kappa^{2}}R-\frac{Z(\Phi)}{4e^{2}}F_{\mu\nu}F^{\mu\nu}
 -\frac{1}{\kappa^{2}}(\partial\Phi)^{2}-\frac{1}{2\kappa^{2}L^{2}}V(\Phi)\right),
\end{equation}
with the effective gauge coupling and scalar potential
\begin{equation}
 Z(\Phi)=Z_{0}^{2}e^{\alpha\Phi},~~~V(\Phi)=-V_{0}^{2}e^{-\beta\Phi}.
\end{equation}
Here $Z_{0}, V_{0}, \alpha, \beta$ are constants characterizing the theory. Such a theory was dubbed as ``Effective Holographic Theory''
in~\cite{Charmousis:2010zz}. It should be emphasized that the following backgrounds with hyperscaling violation and general semi-locality were first investigated in~\cite{Gouteraux:2011ce}.

If the parameters $\alpha$ and $\beta$ take the following values:
\begin{equation}
 \beta=-\frac{\sqrt{8/d}}{1+d/\eta},~~~\alpha=-(d-1)\beta,
\end{equation}
the Einstein-Maxwell-dilaton theory admits the following exact solution at extremality:
\begin{eqnarray}
\label{extIR}
 & &ds^{2}_{d+2}=\frac{L^{2}}{r^{2}}\left(-\frac{dt^{2}}{r^{2d/\eta}}+\frac{g_{0}}{r^{2}}dr^{2}+\sum\limits^{d}_{i=1}dx_{i}^{2}\right),\nonumber\\
 & &g_{0}=\frac{d^{2}}{V_{0}}(1+\frac{1}{\eta})^{2},~~\Phi=\sqrt{\frac{d}{2}}\sqrt{1+\frac{d}{\eta}}\log r,\nonumber\\
 & &A_{t}=\frac{eL}{\kappa}h(r),~~~h(r)=\frac{h_{0}}{r^{d(1+1/\eta)}},~~~h_{0}=\frac{1}{Z_{0}\sqrt{1+\eta}}.
\end{eqnarray}
The IR regime corresponds to taking $r\rightarrow\infty$ while the UV region locates at $r\rightarrow0$.
Such a background possesses the following scaling properties:
\begin{equation}
 t\rightarrow\lambda t, ~~~r\rightarrow\lambda^{\eta/d}r,~\Rightarrow~ds\rightarrow\lambda^{-\eta/d}ds.
\end{equation}
 Furthermore, in this background
only $t$ and $r$ are involved in the scaling while the spatial coordinates $x_{i}$ are spectators, hence the background geometry is ``semi-local''. The finite-temperature counterparts can be written as follows:
\begin{equation}
\label{bh}
 ds^{2}_{d+2}=\frac{L^{2}}{r^{2}}\left(-\frac{f(r)dt^{2}}{r^{2d/\eta}}+\frac{g_{0}}{r^{2}f(r)}dr^{2}+\sum\limits^{d}_{i=1}dx_{i}^{2}\right),
 ~~f(r)=1-(\frac{r}{r_{H}})^{d(1+1/\eta)},
 \end{equation}
 while the other field configurations remain invariant as in the extremal case.

 For simplicity we will consider the following extremal background by setting $g_{0}=1$ in~(\ref{extIR}):
\begin{equation}
 ds^{2}_{d+2}=\frac{L^{2}}{r^{2}}\left(-\frac{dt^{2}}{r^{2d/\eta}}+\frac{dr^{2}}{r^{2}}+\sum\limits_{i=1}^{d}dx_{i}^{2}\right).
\end{equation}
It can be easily seen that upon taking a new radial coordinate $r=\xi^{2/p}$, this metric is conformal to $AdS_{2}\times\mathbb{R}^{d}$:
\begin{equation}
\label{confads2}
ds^{2}=\frac{L^{2}}{\xi^{\frac{2\eta}{d}}}\left[-\frac{dt^{2}}{\xi^{2}}+\frac{d\xi^{2}}{\xi^{2}}+\sum\limits^{d}_{i=1}dx_{i}^{2}\right].
\end{equation}
In addition, the finite-temperature counterpart reads
\begin{equation}
 ds^{2}_{d+2}=\frac{L^{2}}{r^{2}}\left(-\frac{f(r)dt^{2}}{r^{2d/\eta}}+\frac{dr^{2}}{r^{2}f(r)}+\sum\limits_{i=1}^{d}dx_{i}^{2}\right),
 ~~~f(r)=1-\left(\frac{r}{r_{H}}\right)^{d(1+\frac{1}{\eta})}.
\end{equation}
The temperature and entropy density can be easily obtained:
\begin{equation}
 T=\frac{1}{4\pi}d(1+\frac{1}{\eta})r_{H}^{-d/\eta},~~s=\frac{1}{4r_{H}^{d}},
\end{equation}
which leads to a universal scaling behavior $s\sim T^{\eta}$.

\section{Thermodynamics of probe branes}
In this section we discuss thermodynamics of probe D-branes in the backgrounds with semi-locality both at extremality and finite temperature, following~\cite{Karch:2007br}. Before going to the details we emphasize that what we are studying is a toy model, i.e. we are not clear if such backgrounds can be embedded into string theory or can be obtained by superposing D-branes (although certain four- and five-dimensional black hole solutions in supergravity were obtained in~\cite{Cvetic:1999xp}, whose near-horizon geometry is of the type~(\ref{confads2}) with $\eta=1$). However, it is expected that our results may reveal some universal properties in this setup.

In the subsequent discussions we just consider the simplest case: massless probe D-branes with trivial dilaton. Then the DBI action is given by
\begin{equation}
\label{actionDBI}
 S_{\rm DBI}=-N_{f}T_{D}V\int dtdrd^{d}x\sqrt{-{\rm det}(g_{ab}+F_{ab})},
\end{equation}
where $g_{ab}$ and $F_{ab}$ denote the pullback of the metric and the $U(1)$ gauge field strength on the probe D-branes. Notice that we have absorbed the factor $2\pi\alpha^{\prime}$ into $F_{ab}$. Here $N_{f}$ denotes the number of probe branes, $T_{D}$ is the tension of the brane and $V$ is the volume of the internal manifold on which the D-branes wrap.

Let us consider the following general background:
\begin{equation}
 ds^{2}_{d+2}=g_{tt}dt^{2}+g_{rr}dr^{2}+g_{xx}\sum\limits_{i=1}^{d}dx_{i}^{2},~~A=A_{t}(r)dt.
\end{equation}
Substituting the above background into~(\ref{actionDBI}), the DBI action becomes
\begin{equation}
 S_{\rm DBI}=-\mathcal{N}\int drg_{xx}^{d/2}\sqrt{|g_{tt}|g_{rr}-A_{t}^{\prime2}},
\end{equation}
where $\mathcal{N}\equiv N_{f}T_{D}V{\rm Vol}(\mathbb{R}^{1,d})$.
The charge density is given by
\begin{equation}
\label{Jt}
 \langle J^{t}\rangle=\frac{\delta S_{\rm DBI}}{\delta A_{t}^{\prime}}=\frac{\mathcal{N}g_{xx}^{d/2}A_{t}^{\prime}}{|g_{tt}|g_{rr}-A_{t}^{\prime2}}.
\end{equation}
The $U(1)$ gauge field and the on-shell DBI action can be expressed as follows:
\begin{equation}
 A_{t}^{\prime}=\hat{d}\sqrt{\frac{|g_{tt}|g_{rr}}{g_{xx}^{d}+\hat{d}^{2}}},~~S_{\rm DBI}=-\mathcal{N}\int dr
 \frac{g_{xx}^{d}\sqrt{|g_{tt}|g_{rr}}}{g_{xx}^{d}+\hat{d}^{2}},
\end{equation}
where we have introduced $\hat{d}\equiv\langle J^{t}\rangle/\mathcal{N}$.

The chemical potential in the extremal case is obtained by integrating the $U(1)$ gauge field, which gives
\begin{eqnarray}
 \mu_{0}&=&\int^{\infty}_{0}drA^{\prime}_{t}=\hat{d}\int^{\infty}_{0}dr\frac{r^{-(d/\eta+3)}}{\sqrt{\hat{d}^{2}+r^{-2d}}}\nonumber\\
 &=&\frac{\hat{d}^{\frac{1}{\eta}+\frac{2}{d}}}{2d}\frac{\Gamma(\frac{1}{2\eta}+\frac{1}{d})\Gamma(\frac{1}{2}-\frac{1}{2\eta}-\frac{1}{d})}{\Gamma(\frac{1}{2})}.
\end{eqnarray}
Here and in the following the subscript `0' denotes the results obtained in the zero-temperature background.
Moreover, the grand canonical free energy is given by the on-shell action
\begin{eqnarray}
 \Omega_{0}&=&-S_{\rm DBI}=\mathcal{N}\int dr\frac{r^{-(d+d/\eta+3)}}{\sqrt{1+\hat{d}^{2}r^{2d}}}\nonumber\\
 &=&\frac{\mathcal{N}}{2d}\hat{d}^{1+\frac{1}{\eta}+\frac{2}{d}}\frac{\Gamma(1+\frac{1}{2\eta}+\frac{1}{d})
 \Gamma(-\frac{1}{2}-\frac{1}{2\eta-\frac{1}{d}})}{\Gamma(\frac{1}{2})}.
\end{eqnarray}
Hence the charge density reads
\begin{equation}
 \rho=-\frac{\partial\Omega_{0}}{\partial\mu_{0}}=\mathcal{N}\hat{d}.
\end{equation}
Notice that the above result agrees with~(\ref{Jt}), which signifies the consistency of the calculations. The energy density
and the pressure can be obtained following the standard thermodynamical relations:
\begin{equation}
 \epsilon_{0}=\Omega_{0}+\mu_{0}\langle J^{t}\rangle=-\frac{d\eta}{d+2\eta}\Omega_{0},
\end{equation}
\begin{equation}
 P_{0}=-\Omega_{0}=\frac{d+2\eta}{d\eta}\epsilon_{0}.
\end{equation}
Finally the speed of sound reads
\begin{equation}
 v_{0}^{2}=\left(\frac{\partial P_{0}}{\partial\epsilon_{0}}\right)_{\mu}=\frac{1}{\eta}+\frac{2}{d}.
\end{equation}
As emphasized in~\cite{HoyosBadajoz:2010kd}, this quantity is only the speed of normal/first sound in the relativistic case.

Next we consider the finite temperature case, where the chemical potential is still given by integrating the $U(1)$ gauge field:
\begin{eqnarray}
 \mu&=&\int^{r_{H}}_{0}drA_{t}^{\prime}\nonumber\\
 &=&\mu_{0}-\frac{\eta}{d+2\eta}r_{H}^{-(2+d/\eta)}{}_{2}F_{1}[\frac{1}{d}+\frac{1}{2\eta},\frac{1}{2};
 1+\frac{1}{d}+\frac{1}{2\eta}, -\frac{r_{H}^{-2d}}{\hat{d}^{2}}]\nonumber\\
 &=&\mu_{0}-\frac{\eta}{d+2\eta}r_{H}^{-(2+d/\eta)}\left(1-\frac{2\eta+d}{4(d+1)\eta+2d}\frac{r_{H}^{-2d}}{\hat{d}^{2}}
 +\mathcal{O}\left(\frac{r_{H}^{-4d}}{\hat{d}^{2}}\right)\right).
\end{eqnarray}
In this case the grand canonical potential reads
\begin{eqnarray}
 \Omega&=&-S_{\rm DBI}=\mathcal{N}\int^{r_{H}}_{0}dr\frac{r^{-(2d+d/\eta+3)}}{\sqrt{\hat{d}^{2}+r^{-2d}}}\nonumber\\
 &=&\Omega_{0}-\mathcal{N}\frac{r_{H}^{-2(d+1)-d/\eta}}{(2(d+1)+d/\eta)\hat{d}}{}_{2}F_{1}[1+\frac{1}{d}+\frac{1}{2\eta},\frac{1}{2};
 2+\frac{1}{d}+\frac{1}{2\eta}, -\frac{r_{H}^{-2d}}{\hat{d}^{2}}]\nonumber\\
 &=&-\mathcal{N}\hat{d}\frac{d+2\eta}{d(d+2)\eta}\mu-\mathcal{N}\hat{d}\frac{\eta}{d+(d+2)\eta}\left(\frac{4\pi\eta}
 {d(\eta+1)}\right)^{1+2\eta/d}T^{1+2\eta/d}\nonumber\\
 & &+\mathcal{O}(T^{2(1+1/d)\eta+1})
\end{eqnarray}
which leads to the entropy density
\begin{equation}
 s=-\left(\frac{\partial\Omega}{\partial T}\right)_{\mu}\sim T^{2\eta/d}.
\end{equation}
Notice that $s\rightarrow0$ as $T\rightarrow0$, which is quite different from other cases.
It was observed that there is always a nonvanishing entropy density at extremality when the background is Schwarzschild-AdS black hole~\cite{Karch:2008fa} or Lifshitz black hole~\cite{HoyosBadajoz:2010kd}. There the nonzero entropy density at extremality may indicate that we are working with a non-genuine ground state. So here our result shows that such a zero-temperature background may be the genuine ground state.
The specific heat is given by
\begin{equation}
 c_{V}=T\frac{\partial s}{\partial T}\sim T^{2\eta/d},
\end{equation}
Note that the specific heat scales linearly in $T$ for a Fermi liquid and scales as $T^{d}$ for a Bose liquid. Therefore our result matches to the scaling behavior of a Fermi liquid when $d=2\eta$, while it is $d^{2}=2\eta$ for a Bose liquid.

\section{Green's functions at low frequency}
In this section we calculate the retarded Green's functions, from which we will be able to read off the AC conductivity and the zero sound.
The zero sound should appear as a pole in the density-density two-point function at zero temperature and the AC conductivity can be obtained from the current-current two-point function via Kubo's formula. We can calculate the density-density, density-current and current-current two-point functions simultaneously by working with the gauge-invariant variable, following~\cite{HoyosBadajoz:2010kd}.

First let us turn on fluctuations of the gauge fields on the probe D-branes:
\begin{equation}
 A_{\mu}(r)\rightarrow A_{\mu}(r)+a_{\mu}(r,t,x).
\end{equation}
Due to translational invariance in the spatial directions, we may just consider fluctuations $a_{t}, a_{x}$, where $x$ denotes one direction of the $x_{i}$s. After fixing the gauge $a_{r}=0$, we can expand the DBI action up to quadratic order in the fluctuations:
\begin{equation}
 S_{a^{2}}=\frac{\mathcal{N}}{2}\int dtdrdxg_{xx}^{d/2}\left[\frac{g_{rr}f_{tx}^{2}+g_{tt}a_{x}^{\prime2}}{g_{xx}\sqrt{|g_{tt}|g_{rr}-A_{t}^{\prime2}}}+
 \frac{|g_{tt}|g_{rr}a_{t}^{\prime2}}{(|g_{tt}|g_{rr}-A_{t}^{\prime2})^{3/2}}\right],
\end{equation}
where $f_{tx}=\partial_{t}a_{x}-\partial_{x}a_{t}$. Performing a Fourier transformation
\begin{equation}
 a_{\mu}(t,r,x)=\int\frac{d\omega dk}{(2\pi)^{2}}e^{-i\omega t+ikx}a_{\mu}(r,\omega,k),
\end{equation}
the linearized equations of motion are given by
\begin{equation}
\label{eomat}
 \partial_{r}\left[\frac{|g_{tt}|g_{rr}g_{xx}^{d/2}a_{t}^{\prime}}{(|g_{tt}|g_{rr}-A_{t}^{\prime2})^{3/2}}\right]-
 \frac{g_{rr}g_{xx}^{d/2-1}}{\sqrt{|g_{tt}|g_{rr}}-A_{t}^{\prime2}}(k^{2}a_{t}+\omega ka_{x})=0,
\end{equation}
\begin{equation}
\label{eomax}
 \partial_{r}\left[\frac{|g_{tt}|g_{xx}^{d/2-1}a_{x}^{\prime}}{\sqrt{|g_{tt}|g_{rr}-A_{t}^{\prime2}}}\right]+
 \frac{g_{rr}g_{xx}^{d/2-1}}{\sqrt{|g_{tt}|g_{rr}}-A_{t}^{\prime2}}(\omega^{2}a_{x}+\omega ka_{t})=0.
\end{equation}
Moreover, we have another constraint which comes from the equation of motion of $a_{r}$ by setting $a_{r}=0$:
\begin{equation}
\label{eomar}
 g_{rr}g_{xx}\omega a_{t}^{\prime}+(|g_{tt}|g_{rr}-A_{t}^{\prime2})ka_{x}^{\prime}=0,
\end{equation}
Indeed the above three equations~(\ref{eomat}), (\ref{eomax}), (\ref{eomar}) are not totally independent. It can be seen that the equation of motion for $a_{t}$ can be reproduced by solving for $a_{x}^{\prime}$ from~(\ref{eomar}) and plugging the result into~(\ref{eomax}). Therefore we only need to solve the constraint and the equation of motion for $a_{t}$.

In the following we will work with the gauge invariant variable $E(r,\omega,k)=\omega a_{x}+ka_{t}$,
which results in the following equation of motion for $E$:
\begin{equation}
\label{eomE}
 E^{\prime\prime}+\left[\partial_{r}\ln\left(\frac{|g_{tt}|g_{rr}^{-1/2}g_{xx}^{(d-3)/2}}{u(k^{2}u^{2}-\omega^{2})}\right)\right]
 -\frac{g_{rr}}{|g_{tt}|}(k^{2}u^{2}-\omega^{2})E=0,
\end{equation}
where
\begin{equation}
 u(r)=\sqrt{\frac{|g_{tt}|g_{rr}-A_{t}^{\prime2}}{g_{rr}g_{xx}}}.
\end{equation}
We can also rewrite the quadratic DBI action in terms of $E$:
\begin{equation}
 S_{a^{2}}=\frac{\mathcal{N}}{2}\int drd\omega dk\frac{g_{xx}^{(d-3)/2}g_{rr}^{1/2}}{u}\left[E^{2}
 +\frac{|g_{tt}|}{g_{rr}(u^{2}k^{2}-\omega^{2})}E^{\prime2}\right].
\end{equation}
The on-shell boundary action can be obtained by introducing a cutoff at $r=\epsilon$, integrating by parts and taking the $\epsilon\rightarrow0$ limit:
\begin{equation}
 S_{a^{2}}=-\frac{N}{2}\int d\omega dk\frac{\epsilon^{d/\eta-d+3}}{k^{2}}E(\epsilon)E^{\prime}(\epsilon).
\end{equation}
To calculate the Green's functions, we need to solve the equation of motion with the ingoing boundary condition at $r\rightarrow\infty$. Then we substitute the solution into $S_{a^{2}}$ and functionally differentiate, e.g.,
\begin{equation}
G^{R}_{tt}(\omega, k)=\frac{\delta^{2}}{\delta a_{t}(\epsilon)^{2}}S_{a^{2}}=\left(\frac{\delta E(\epsilon)}{\delta a_{t}(\epsilon)}\right)^{2}\frac{\delta^{2}}{\delta E(\epsilon)^{2}}S_{a^{2}}.
\end{equation}
If we define the following quantity
\begin{equation}
 \Pi(\omega,k)\equiv\frac{\delta^{2}}{\delta E(\epsilon)^{2}}S_{a^{2}},
\end{equation}
the retarded Green's functions can be expressed in terms of $\Pi(\omega,k)$:
\begin{equation}
 G^{R}_{tt}(\omega,k)=k^{2}\Pi(\omega,k),~~G^{R}_{tx}(\omega,k)=\omega k\Pi(\omega,k),~~G^{R}_{xx}(\omega,k)=\omega^{2}\Pi(\omega,k).
\end{equation}
Therefore our task is to work out $\Pi(\omega, k)$.

Generally, it is very difficult to find analytic solutions to the equation of motion for $E(u)$~(\ref{eomE}). However, if we just care about the low-frequency behavior, we can solve~(\ref{eomE}) in two different limits and then match the two solutions in certain regime where the two limits overlap~\cite{Karch:2008fa, HoyosBadajoz:2010kd}. For our particular case, the first solution can be obtained by taking the large $r$, small frequency and momentum limit, which means $\omega r^{d/\eta}\ll1, kr\ll1$, with $\omega k^{-d/\eta}$ fixed. The second solution can be obtained in a similar way, but with taking the small frequency and momentum limit first. Now the equation of motion for $E$ in the $r\rightarrow\infty$ limit reads
\begin{equation}
 E^{\prime\prime}+\frac{1}{\eta}(3-\frac{d}{\eta})E^{\prime}+\omega^{2}r^{2d/\eta-2}E=0.
\end{equation}
The solution can be expressed in terms of the Hankel function of the first kind:
\begin{equation}
 E=C\left(\frac{\omega r^{d/\eta}}{2d/\eta}\right)^{\frac{1}{2}-\frac{\eta}{d}}H^{(1)}_{\frac{1}{2}-\frac{\eta}{d}}(\frac{\omega r^{d/\eta}}{d/\eta}).
\end{equation}
Since the solution does not involve the momentum $k$, we can just perform the small frequency expansion. However, the expansion does depend on the value of $d/\eta$. When $\eta/d\neq1/2$, the small frequency expansion is given by
\begin{equation}
 E\simeq C\Gamma(\frac{\eta}{d}+\frac{1}{2})^{-1}(1-i\tan\frac{\pi}{2})-i\frac{C}{\pi}\Gamma(\frac{\eta}{d}-\frac{1}{2})(\frac{\omega\eta}{2d})^{1-2\eta/d}r^{d/\eta-2},
\end{equation}
while a logarithmic term appears when $\eta/d=1/2$:
\begin{equation}
 E\simeq C+C\frac{2i}{\pi}(\log\omega r^{2}-\log4+\gamma),
\end{equation}
where $\gamma$ denotes the Euler-Mascheroni number.

Next after taking the limit $\omega r^{d/\eta}\ll1, kr\ll1$ with $\omega k^{-d/\eta}$ fixed, the equation of motion for $E$ becomes
\begin{equation}
 E^{\prime\prime}+\left(\frac{3-d-2d/\eta}{r}-\frac{u^{\prime}}{u}\frac{3k^{2}u^{2}-\omega^{2}}{k^{2}u^{2}-\omega^{2}}\right)E^{\prime}=0.
\end{equation}
The solution can be expressed in terms of the hypergeometric function
\begin{eqnarray}
 E&=&C_{1}+C_{2}\eta r^{d-d/\eta-2}\Big\{\frac{\omega^{2}r^{2d/\eta}}{d-2\eta+d\eta}{}_{2}F_{1}[\frac{3}{2},\frac{1}{2}-\frac{1}{d}+\frac{1}{2\eta},
 \frac{3}{2}-\frac{1}{d}+\frac{1}{2\eta},-\hat{d}^{2}r^{2d}]\nonumber\\
 & &+\frac{\hat{d}^{2}\omega^{2}r^{2d/\eta+2d}}{d-2\eta+3d\eta}{}_{2}F_{1}[\frac{3}{2},\frac{3}{2}-\frac{1}{d}+\frac{1}{2\eta},
 \frac{5}{2}-\frac{1}{d}+\frac{1}{2\eta},-\hat{d}^{2}r^{2d}]\nonumber\\
 & &-\frac{k^{2}}{d\eta-2\eta-d}{}_{2}F_{1}[\frac{3}{2},\frac{1}{2}-\frac{1}{d}-\frac{1}{2\eta},
 \frac{3}{2}-\frac{1}{d}-\frac{1}{2\eta},-\hat{d}^{2}r^{2d}]\Big\}.
\end{eqnarray}
In order to match the second solution with the first solution, we shall make use of the following formulae:
\begin{eqnarray}
 {}_{2}F_{1}[a,b;c,z]&=&\frac{\Gamma(c)\Gamma(b-a)}{\Gamma(b)\Gamma(c-a)}(-z)^{-a}{}_{2}F_{1}[a,1-c+a;1-b+a;\frac{1}{z}]\nonumber\\
 & &+\frac{\Gamma(c)\Gamma(a-b)}{\Gamma(a)\Gamma(c-b)}(-z)^{-b}{}_{2}F_{1}[b,1-c+b;1-a+b;\frac{1}{z}]\nonumber\\
 {}_{2}F_{1}[a,b;c,z]&\simeq&1~~{\rm as}~z\rightarrow0.
\end{eqnarray}
Thus when $d/\eta\neq2$, the large $r$ expansion is given by
\begin{eqnarray}
 E&\simeq& C_{1}+C_{2}k^{2}\hat{d}^{1/\eta+2/d-1}(\frac{1}{d^{2}}+\frac{1}{2d\eta})B(\frac{1}{2}-\frac{1}{2\eta}-\frac{1}{d},\frac{1}{2\eta}+\frac{1}{d})\nonumber\\
 & &-C_{2}\frac{\omega^{2}r^{d/\eta-2}}{\hat{d}(d/\eta-2)}-C_{2}\frac{\omega^{2}\hat{d}^{2/d-1/\eta-1}}{2d}B(\frac{1}{2}-\frac{1}{d}+\frac{1}{2\eta},
 \frac{1}{d}-\frac{1}{2\eta}),
\end{eqnarray}
while we have the following result when $d/\eta=2$:
\begin{equation}
 E\simeq C_{1}+C_{2}k^{2}\hat{d}^{4/d-1}\frac{2}{d^{2}}B(\frac{1}{2}-\frac{2}{d},\frac{2}{d})-C_{2}\frac{\omega^{2}}{d\hat{d}}\log(2\hat{d}r^{d}).
\end{equation}
In addition, we rewrite the following term for later convenience:
\begin{equation}
 \log(2\hat{d}r^{d})=\log(\frac{2\hat{d}}{\omega^{d/2}})+\frac{d}{2}\log(\omega r^{2}).
\end{equation}

Notice that in the limit of small $r$, the hypergeometric function $\sim1$ and the $k^{2}$ term is dominant, so the small $r$ expansion is given by
\begin{equation}
 E\simeq C_{1}+C_{2}\frac{k^{2}r^{d-d/\eta-2}}{d-d/\eta-2}.
\end{equation}
Let us focus on the case $d-d/\eta-2>0$, as the other cases can be deduced in a similar way. For details see footnote 8 of~\cite{HoyosBadajoz:2010kd}.
We have $E(\epsilon)\simeq C_{1}$ to leading order, so
\begin{equation}
 \Pi(\omega,k)=\lim_{\epsilon0}\frac{\delta^{2}}{\delta E(\epsilon)^{2}}S_{a^{2}}
 =\frac{\delta^{2}}{\delta C_{1}^{2}}S_{a^{2}}\big|_{\epsilon\rightarrow0}.
\end{equation}
Moreover, the leading order expansion of $E^{\prime}$ reads
\begin{equation}
 E^{\prime}(\epsilon)\simeq C_{2}k^{2}\epsilon^{d-d/\eta-3},
\end{equation}
so
\begin{equation}
 S_{a^{2}}=-\frac{\mathcal{N}}{2}\int d\omega dk\frac{\epsilon^{d/\eta-d+3}}{k^{2}}E(\epsilon)E^{\prime}(\epsilon)=
 -\frac{\mathcal{N}}{2}\int d\omega dk C_{1}C_{2}.
\end{equation}
To perform the matching, we should first match the constant terms, which will result in an expression of $C$ in terms of $C_{2}$. Then we can read off $C_{2}$ in terms of $C_{1}$ by matching the $r$-dependent terms. Finally we obtain
\begin{equation}
\label{Pi}
 \Pi(\omega,k)\propto\frac{\mathcal{N}}{\alpha_{1}k^{2}-\alpha_{2}\omega^{2}-\alpha_{3}G_{0}(\omega)},
\end{equation}
where
\begin{eqnarray}
 & &\alpha_{1}=\frac{(2-\frac{d}{\eta})\Gamma(\frac{\eta}{d}-\frac{1}{2})\Gamma(\frac{\eta}{d}+\frac{1}{2})}{\pi(2d/\eta)^{1-2\eta/d}}\hat{d}^{\frac{1}{\eta}
 +\frac{2}{d}}\left(\frac{1}{d^{2}}+\frac{1}{2d\eta}\right)B(\frac{1}{2}-\frac{1}{2\eta}-\frac{1}{d},\frac{1}{2\eta}+\frac{1}{d}),\nonumber\\
 & &\alpha_{2}=\frac{(2-\frac{d}{\eta})\Gamma(\frac{\eta}{d}-\frac{1}{2})\Gamma(\frac{\eta}{d}+\frac{1}{2})}{\pi(2d/\eta)^{1-2\eta/d}}\hat{d}^{
 \frac{2}{d}-\frac{1}{\eta}}B(\frac{1}{2}+\frac{1}{2\eta}-\frac{1}{d},\frac{1}{d}-\frac{1}{2\eta}),\nonumber\\
 & &\alpha_{3}=i+\tan\frac{\pi\eta}{d},~~~G_{0}(\omega)=\omega^{1+2\eta/d}.
\end{eqnarray}
The $d/\eta=2$ case should be treated separately, but the resulting structure is the same as~(\ref{Pi}), with
\begin{eqnarray}
 & &\alpha_{1}=\frac{8\hat{d}^{4/d}}{\pi d^{2}}B(\frac{1}{2}-\frac{2}{d},\frac{2}{d}),~~\alpha_{2}=i,~~\alpha_{3}=-\frac{1}{\pi},\nonumber\\
 & &G_{0}(\omega)=\omega^{2}\log\alpha\omega^{2},~~\alpha=\frac{e^{2\gamma}}{16(2\hat{d})^{4/d}}.
\end{eqnarray}
Once we have obtained $\Pi(\omega,k)$ we may calculate the AC conductivity and the zero sound.
\section{The AC conductivity}
The AC conductivity can be read off from the current-current correlation function by Kubo's formula:
\begin{equation}
 \sigma(\omega)=-\frac{i}{\omega}G^{R}_{xx}(\omega,k=0)=-i\omega\Pi(\omega,k=0).
\end{equation}
Notice that when $d/\eta\neq2$,
\begin{equation}
 -i\omega\Pi(\omega,k=0)\simeq\frac{i\mathcal{N}}{\alpha_{2}\omega+\alpha_{3}\omega^{2\eta/d}},
\end{equation}
therefore the AC conductivity is given by
\begin{eqnarray}
 & &\sigma(\omega)\sim\frac{i\mathcal{N}}{\alpha_{2}\omega},~~d/\eta<2,\nonumber\\
 & &\sigma(\omega)\sim\frac{i\mathcal{N}}{\alpha_{3}\omega^{2\eta/d}},~~d/\eta>2.
\end{eqnarray}
When $d/\eta=2$, the AC conductivity contains a logarithmic term:
\begin{equation}
 \sigma(\omega)\sim\frac{i\mathcal{N}}{\alpha_{3}\omega\log(\alpha\omega^{2})}.
\end{equation}

Notice that when $d/\eta\neq2$, $\alpha_{2}$ is a real number and $\alpha_{3}$ is complex. When $d/\eta<2$, the conductivity is purely imaginary and has a simple pole at zero frequency. Therefore the real part of the conductivity consists only of a delta function at zero frequency, according to the Kramers-Kronig relation, so does the spectral function. However, when $d/\eta>2$ the conductivity and hence the spectral function have power-law dependence. In the next section we will see that holographic zero sound mode can exist for $d/\eta<2$, while there is no holographic zero sound quasi-particle for $d/\eta>2$.
\section{The zero sound}
In this section we calculate the holographic zero sound. Such a sound mode can be read off from the poles of the density-density retarded Green's function. The holographic zero sound mode from a probe brane setup was discussed in~\cite{Karch:2008fa} where the background was $AdS$ spacetime.  Subsequently this formalism was generalized to D3/D7 branes with finite mass in~\cite{Kulaxizi:2008kv}, and to Sakai-Sugimoto model in~\cite{Kulaxizi:2008jx}. Moreover, the zero sound in Lifshitz spacetime was discussed in~\cite{HoyosBadajoz:2010kd} and the zero sound in background with hyperscaling violation was studied in~\cite{Lee:2010ez}. Generalizations to finite temperature cases have been carried out in~\cite{Davison:2011ek, Davison:2011uk}.

As discussed above, the holographic zero sound can be read off from the poles of $\Pi(\omega,k)$, which results in
\begin{equation}
 k(\omega)=\pm\frac{1}{\sqrt{\alpha_{1}}}\sqrt{\alpha_{2}\omega^{2}+\alpha_{3}G_{0}(\omega)}.
\end{equation}
Notice that when $d/\eta\neq2$, the above expression becomes
\begin{equation}
 k(\omega)=\pm\frac{1}{\sqrt{\alpha_{1}}}\sqrt{\alpha_{2}\omega^{2}+\alpha_{3}\omega^{1+2\eta/d}}.
\end{equation}
In order to classify the behavior, we should discuss the cases $d/\eta<2$ and $d/\eta>2$ separately. Notice that when $d/\eta<2$, the $\omega^{2}$ term dominates:
\begin{eqnarray}
 k(\omega)&=&\pm\omega\sqrt{\frac{\alpha_{2}}{\alpha_{1}}}\sqrt{1+\frac{\alpha_{3}}{\alpha_{2}}\omega^{-1+2\eta/d}}\nonumber\\
 &=&\pm\omega\sqrt{\frac{\alpha_{2}}{\alpha_{1}}}\left[1+\frac{\alpha_{3}}{2\alpha_{2}}\omega^{-1+2\eta/d}+\mathcal{O}(\omega^{-2+4\eta/d})\right].
\end{eqnarray}
We can invert the above expression to find
\begin{equation}
 \omega(k)=\pm k\sqrt{\frac{\alpha_{1}}{\alpha_{2}}}-\frac{\alpha_{3}}{2\alpha_{2}}\left(\frac{\alpha_{1}}{\alpha_{2}}\right)^{\eta/d}k^{2\eta/d}+
 \mathcal{O}(k^{-1+4\eta/d}).
\end{equation}
However, more classifications are required when $d$ and $\eta$ take different values:
\begin{itemize}
 \item When $d=2$, we can obtain $\eta>1$, $\alpha_{1}<0$. It can be seen that the coefficient of the first term is imaginary, which does not correspond a quasi-particle.
 \item When $d=3$, we have $\eta>3/2$ and we can see that when $3/2<\eta<3$, $\alpha_{1}<0$, which does not correspond to a quasi-particle. Moreover,  $\alpha_{1}$ diverges when $\eta=3$.  Only when $\eta>3$ we can find quasi-particle with zero sound:
 \begin{equation}
  v_{0}^{2}=\frac{\alpha_{1}}{\alpha_{2}}=\hat{d}^{2/\eta}(\frac{1}{9}+\frac{1}{6\eta})\frac{\Gamma(\frac{1}{6}-\frac{1}{2\eta})\Gamma(\frac{1}{2\eta}+\frac{1}{3})}
  {\Gamma(\frac{1}{6}+\frac{1}{2\eta})\Gamma(\frac{1}{3}-\frac{1}{2\eta})},
 \end{equation}
\item When $d=4$, we can obtain quasi-particle description with constant zero sound:
 \begin{equation}
  v_{0}^{2}=\frac{\alpha_{1}}{\alpha_{2}}=\hat{d}^{2/\eta}(\frac{1}{16}+\frac{1}{8\eta}).
 \end{equation}
\item When $d\geq5$, $\alpha_{1},\alpha_{2}$ are always positive. Therefore we can find quasi-particle with zero sound:
\begin{equation}
 v_{0}^{2}=\frac{\alpha_{1}}{\alpha_{2}}=\hat{d}^{2/\eta}(\frac{1}{d^{2}}+\frac{1}{2d\eta})
 \frac{\Gamma(\frac{1}{2}-\frac{1}{2\eta}-\frac{1}{d})\Gamma(\frac{1}{2\eta}+\frac{1}{d})}
 {\Gamma(\frac{1}{2}-\frac{1}{d}+\frac{1}{2\eta})\Gamma(\frac{1}{d}-\frac{1}{2\eta})},
\end{equation}
\end{itemize}

Next we consider the case $d/\eta>2$, which leads to
\begin{eqnarray}
 k(\omega)&=&\pm\sqrt{\frac{\alpha_{3}}{\alpha_{1}}}\omega^{1/2+\eta/d}\left(1+\frac{\alpha_{2}}{\alpha_{3}}\omega^{1-2\eta/d}\right)^{1/2}\nonumber\\
 &=&\pm\sqrt{\frac{\alpha_{3}}{\alpha_{1}}}\omega^{1/2+\eta/d}\left(1+\frac{\alpha_{2}}{2\alpha_{3}}\omega^{1-2\eta/d}+\mathcal{O}(\omega^{2-4\eta/d})\right).
\end{eqnarray}
Inverting the above result we get
\begin{equation}
 \omega(k)=\left(\frac{\alpha_{1}}{i\alpha_{3}}\right)^{\frac{d}{d+2\eta}}k^{\frac{2d}{d+2\eta}}+\frac{\alpha_{2}}{\alpha_{3}}
 \frac{d}{d+2\eta}\left(\frac{\alpha_{1}}{\alpha_{3}}\right)^{\frac{2(d-\eta)}{d+2\eta}}k^{\frac{4(d-\eta)}{d+2\eta}}+\mathcal{O}(k^{\frac{2(3d-4\eta)}{d+2\eta}}).
\end{equation}
Note that the leading term has a complex coefficient, which means that the real and imaginary parts are of the same order. Therefore the excitation is not a quasi-particle.

Finally we come to the case $d/\eta=2$:
\begin{equation}
 k=\pm\frac{\omega}{\sqrt{\alpha_{1}}}\sqrt{\alpha_{2}+\alpha_{3}\log(\alpha\omega^{2})}.
\end{equation}
Expanding for small $\omega$ we find
\begin{equation}
 k(\omega)=\pm\frac{\omega}{\sqrt{\alpha_{1}}}\sqrt{\alpha_{3}\log(\alpha\omega^{2})}-\frac{\omega}{\sqrt{\alpha_{1}}}
 \frac{\alpha_{2}}{2}(\alpha_{3}\log(\alpha\omega^{2}))^{-1/2}+\mathcal{O}(\omega\log^{-3/2}(\alpha\omega^{2})).
\end{equation}
Although this mode is gapless, the dispersion relation differs from the holographic zero sound mode due to the logarithmic factors.
\section{Summary and discussion}
We study dynamics of probe D-branes in backgrounds with general semi-locality, which are conformal to $AdS_{2}\times\mathbb{R}^{d}$. Thermodynamical quantities are calculated and we find that the entropy density is vanishing in the extremal limit, which is desirable compared to the cases studied in e.g.~\cite{Karch:2008fa, HoyosBadajoz:2010kd} where a nonzero entropy density at extremality was observed. This result indicates that the extremal background we study may be the true ground state of the system. The specific heat is also calculated and the conditions under which the behavior of the specific heat matches to a Bose liquid or a Fermi liquid are clarified. We compute the retarded density-density and current-current correlation functions by working with the gauge-invariant quantity. The AC conductivity and zero sound mode can be obtained from the current-current and density-density Green's functions respectively. We find that the AC conductivity exhibits different scaling behavior, depending on the value of $d/\eta$. When $d/\eta<2$, the AC conductivity scales as $\sigma(\omega)\sim\omega^{-1}$, while $\sigma(\omega)\sim\omega^{-2\eta/d}$ when $d/\eta>2$. There is a logarithmic term in $\omega$ in $\sigma(\omega)$ at the critical value $d/\eta=2$. In particular, when $d/\eta=3$, the AC conductivity matches to the behavior of strange metal~\cite{Hartnoll:2009ns}. Generally, the behavior of the AC conductivity is analogous to the case studied in~\cite{HoyosBadajoz:2010kd}, where the background is Lifshitz black hole and the critical value of the parameter is $z=2$.

In~\cite{HoyosBadajoz:2010kd} it was observed that the behaviors of the AC conductivity and the zero sound mode have deeper relations. When
$z>2$ the conductivity and the spectral function have  power-law dependence, and no clean holographic zero sound quasi-particle exists, while a quasi-particle excitation exists when $z<2$. In our case we find that there is no quasi-particle description for $d/\eta>2$, which is analogous to the $z>2$ case. However, when $d/\eta<2$ it is not necessarily guaranteed that there exists quasi-particle and we study all the possible situations case by case.

\bigskip \goodbreak \centerline{\bf Acknowledgments}
\noindent
DWP is supported by Alexander von Humboldt Foundation.

\newpage

\end{document}